\documentclass[aps,prl,reprint,superscriptaddress]{revtex4-1}
\usepackage{times}
\usepackage{graphicx}
\usepackage{bbm}
\usepackage{siunitx} 
\usepackage{amssymb}
\usepackage{comment}
\usepackage{amsmath} 
\usepackage[hidelinks]{hyperref} 

\newcommand{\fignum}[1]{(\textbf{Fig.~#1})}
\newcommand{\fignumnp}[1]{\textbf{Fig.~#1}}
\newcommand{\ket}[1]{\left| #1 \right>} 

\newcounter{lastnote}

\begin{document}

\title{
Photon-mediated interactions between quantum emitters\\ in a diamond nanocavity 
}


\author{Ruffin E. Evans}
\thanks{These authors contributed equally.}
\affiliation{Department of Physics, Harvard University, 17 Oxford St., Cambridge, MA 02138}
\author{Mihir K. Bhaskar}
\thanks{These authors contributed equally.}
\affiliation{Department of Physics, Harvard University, 17 Oxford St., Cambridge, MA 02138}
\author{Denis D. Sukachev}
\thanks{These authors contributed equally.}
\affiliation{Department of Physics, Harvard University, 17 Oxford St., Cambridge, MA 02138}
\author{Christian T. Nguyen}
\affiliation{Department of Physics, Harvard University, 17 Oxford St., Cambridge, MA 02138}
\author{Alp Sipahigil}
\affiliation{Department of Physics, Harvard University, 17 Oxford St., Cambridge, MA 02138}
\author{Michael J. Burek}
\affiliation{John A. Paulson School of Engineering and Applied Sciences, 29 Oxford St., Cambridge, MA 02138}
\author{Bartholomeus Machielse}
\affiliation{Department of Physics, Harvard University, 17 Oxford St., Cambridge, MA 02138}
\affiliation{John A. Paulson School of Engineering and Applied Sciences, 29 Oxford St., Cambridge, MA 02138}
\author{Grace H. Zhang}
\affiliation{Department of Physics, Harvard University, 17 Oxford St., Cambridge, MA 02138}
\author{Alexander S. Zibrov}
\affiliation{Department of Physics, Harvard University, 17 Oxford St., Cambridge, MA 02138}
\author{Edward Bielejec}
\affiliation{Sandia National Laboratories, Albuquerque, NM 87185, USA}
\author{Hongkun Park}
\affiliation{Department of Physics, Harvard University, 17 Oxford St., Cambridge, MA 02138}
\affiliation{Department of Chemistry and Chemical Biology, Harvard University, 12 Oxford St., Cambridge, MA 02138}
\author{Marko Lon\v{c}ar}
\affiliation{John A. Paulson School of Engineering and Applied Sciences, 29 Oxford St., Cambridge, MA 02138}
\author{Mikhail D. Lukin}
\thanks{lukin@physics.harvard.edu}
\affiliation{Department of Physics, Harvard University, 17 Oxford St., Cambridge, MA 02138}

\begin{abstract}
Photon-mediated interactions between quantum systems are essential for realizing quantum networks and scalable quantum information processing.
We demonstrate such interactions
between pairs of silicon-vacancy (SiV) color centers strongly coupled to a diamond nanophotonic cavity.
When the optical transitions of the two color centers are tuned into resonance,
the coupling to the common cavity mode results in a coherent interaction between them, 
leading to spectrally-resolved superradiant and subradiant states.
We use the electronic spin degrees of freedom of the SiV centers to control these optically-mediated 
interactions.
Our experiments
pave the way for implementation of  
cavity-mediated quantum gates between spin qubits and for realization of 
scalable quantum network nodes.
\end{abstract}

\maketitle

Photon-mediated interactions between quantum emitters are an important building block of many quantum information systems,
enabling entanglement generation and quantum logic operations involving both stationary quantum bits (qubits) and photons \cite{kimble2008quantum,imamoglu1999quantum}.
Recent progress in the field of cavity quantum electrodynamics (QED)
with trapped atoms and ions \cite{reiserer2015cavity},
spin qubits in silicon \cite{samkharadze2018strong,mi2018coherent}, 
superconducting qubits \cite{wallraff2004strong} and
self-assembled quantum dots \cite{lodahl2015interfacing}
has opened up several avenues for engineering and controlling such interactions. 
In particular, coherent interactions between multiple qubits mediated via a cavity mode have been demonstrated in the microwave domain using circuit QED \cite{majer2007coupling}. This technique is now an essential element of superconducting quantum processors \cite{dicarlo2009demonstration}.
Extending such coherent interactions into the optical domain could
enable the implementation of key protocols in long-distance quantum communication and quantum networking \cite{cirac1997quantum,monroe2014large,kastoryano2011dissipative}, dramatically increasing the speed\cite{borregaard2015long} of existing optical entanglement protocols \cite{humphreys2018deterministic} and the communication distance by obviating the need for a cryogenic bus required for microwave photons \cite{kurpiers2018deterministic}.
This goal is challenging due to the difficulty of achieving strong cavity coupling and individual control of multiple resonant quantum emitters.
Recently, two-ion \cite{casabone2015enhanced} and two-atom \cite{reimann2015cavity,neuzner2016interference} systems have been 
used to observe cavity-modified collective scattering.
Spectral signatures of cavity-mediated interactions between quantum dots have also been reported \cite{laucht2010mutual,kim2011strong}.
However, the realization of controlled, coherent optical interactions between solid-state emitters is particularly difficult due to inhomogeneous broadening and decoherence introduced by the solid-state environment \cite{kim2011strong,lodahl2015interfacing}.

\begin{figure*}[t]
\begin{center}
		\includegraphics[width=1.75\columnwidth]{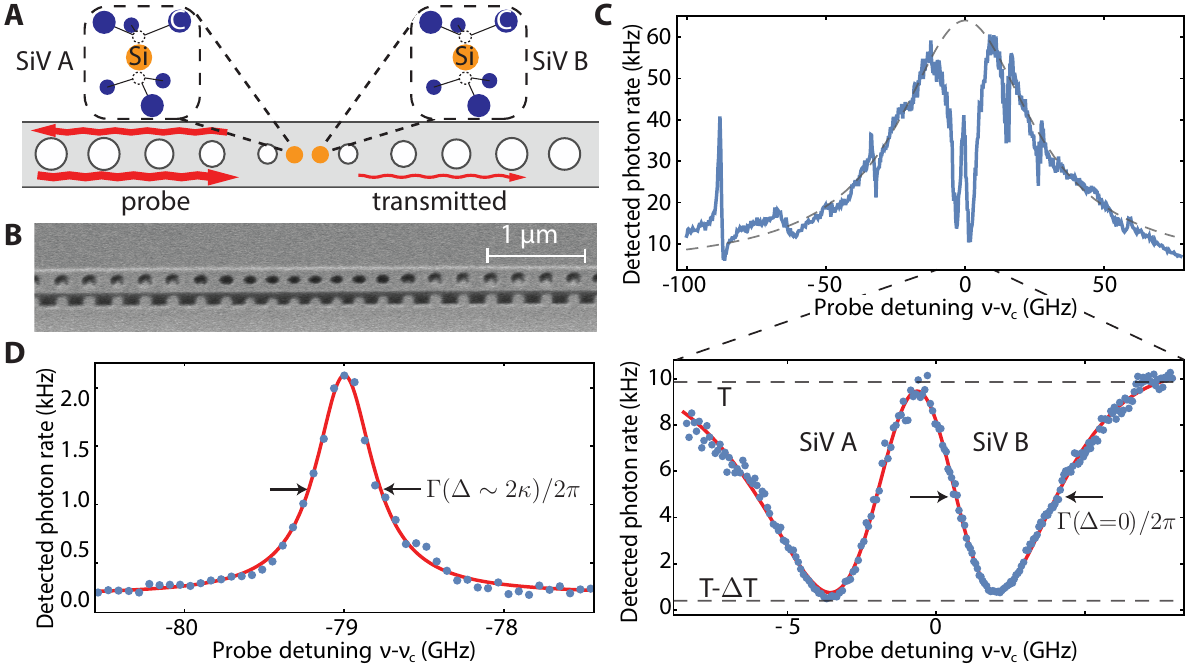}
\end{center}
		\vspace{-0.25cm}
		\caption{
	\textbf{High cooperativity SiV-photon interface.}
	\textbf{(A)} 
	Multiple SiVs are deterministically positioned in a nanophotonic cavity which is probed in transmission.
	\textbf{(B)}
	Scanning electron micrograph of a diamond nanophotonic cavity.
	\textbf{(C)}
	Transmission spectrum of the coupled SiV-cavity system (blue). The broad Lorentzian response of an empty cavity (dashed) is modulated by SiVs coupled to the cavity.
	Near the cavity resonance (lower panel), two SiVs each result in greater than 95\% extinction in transmission and are broadened by the Purcell effect ($\Gamma(\Delta=0) =\,2\pi\times$\SI{4.6}{GHz}).
	\textbf{(D)}
	Far detuned from the cavity resonance ($\Delta =\,2\pi\times$\SI{79}{GHz}\ $\sim2\kappa$),
	individual SiVs appear as narrow peaks in transmission ($\Gamma(\Delta) =\,2\pi\times$\SI{0.5}{GHz}). 
	The solid lines in (D) and the lower panel of (C) are fits to a model \cite{SOM}. 
		}
	\label{fig:system}
\end{figure*}

We realize controllable optically-mediated interactions between negatively-charged silicon-vacancy (SiV) color centers coupled to a diamond photonic crystal cavity \fignum{1A} \cite{sipahigil2016integrated,burek2014high}.
The SiV center in diamond is an atom-like quantum emitter featuring nearly lifetime-limited optical linewidths with low inhomogeneous broadening, both in bulk \cite{sipahigil2014indistinguishable} and, crucially, in nanostructures \cite{evans2016narrow}. We integrate SiV centers into devices consisting of a one-dimensional freestanding diamond waveguide with an array of holes defining a photonic crystal cavity with quality factor $Q\sim{10^4}$ and simulated mode volume $V\sim0.5\left(\frac{\lambda}{n=2.4}\right)^3$ \fignum{1B} \cite{burek2017fiber}.
SiV centers are positioned in these devices with \SI{40}{nm} precision by targeted implantation of $^{29}\mathrm{Si}$ ions using a focused ion beam, yielding around 5 SiV centers per device \cite{sipahigil2016integrated}. One end of the diamond waveguide is tapered and adiabatically coupled to a tapered single-mode fiber, enabling collection efficiencies from the waveguide into the fiber of more than 90\% \cite{burek2017fiber}.
These devices are placed in a dilution refrigerator with an integrated confocal microscope \cite{SOM}.
By working at \SI{85}{mK}, we completely polarize the SiV centers into the lowest-energy orbital state \cite{sukachev2017silicon} used in this work.
We also apply an up to \SI{10}{kG} magnetic field to lift the SiV center's electronic spin degeneracy \cite{sukachev2017silicon,hepp2014electronic}.
To control the SiV-cavity detuning ($\Delta = \omega_{c}-\omega_{\mathrm{SiV}}$), we tune the cavity resonance frequency $\omega_c$ using gas condensation \cite{SOM}. 

The coupling between SiV centers and the nanophotonic cavity is characterized by scanning the frequency of an excitation laser incident on one side of the device from free space while monitoring the intensity of the transmitted field into the collection fiber.
The resulting transmission spectrum of the SiV-cavity system (\fignumnp{1C}, upper spectrum) reveals strong modulation of the bare cavity response resulting from the coupling of multiple spectrally-resolved SiV centers to the cavity mode.
For instance, two SiV centers near the cavity resonance each result in almost-full extinction of the transmission through the cavity ($\Delta T/T > 95\%$; \fignumnp{1C}, lower spectrum) \cite{waks2006dipole}.
In contrast, when the cavity is detuned from the SiV by several cavity linewidths ($\kappa$), the transmission spectrum shows a narrow peak near each SiV frequency \fignum{1D}.
This resonance corresponds to an atom-like dressed state of the strongly-coupled SiV-cavity system featuring a high transmission amplitude \cite{majer2007coupling,faraon2008dipole}.
We also observe that the resonance linewidth ($\Gamma$) changes by more than an order of magnitude depending on the SiV-cavity detuning ($\Delta$).
This observation can be understood in terms of Purcell enhancement, which predicts
$\Gamma(\Delta)\approx\gamma+\frac{4g^2}{\kappa}\frac{1}{1+4\Delta^2/\kappa^2}$ where $g$ is the single-photon Rabi frequency, $\kappa$ is the cavity energy decay rate and $\gamma$ is twice the decoherence rate due to free-space spontaneous emission and spectral diffusion.
For the most strongly-coupled SiV in the device used in \fignumnp{1}, we measure linewidths ranging from $\Gamma(0) = 2\pi \times4.6$ \si{GHz} on resonance to $\Gamma(7 \kappa) = 2\pi \times 0.19$ \si{GHz} $\approx\gamma$ when the cavity is far detuned.
The measured $\Gamma(0)$ corresponds to an estimated Purcell-reduced lifetime of \SI{35}{ps} compared to the natural SiV lifetime of \SI{1.8}{ns} \cite{sipahigil2016integrated}.
From these measurements, we extract the cavity QED 
parameters $\{g,\kappa,\gamma\}=2\pi \times \{7.3,48,0.19\}$ \si{GHz}.
These parameters correspond to a cooperativity (the key cavity-QED figure of merit) of $C=\frac{4 g^2}{\kappa \gamma} \sim 23$
 \cite{SOM}.
This order-of-magnitude improvement in SiV-cavity cooperativity over previous work \cite{sipahigil2016integrated,zhang2018strongly} 
primarily results from the decreased mode volume of the cavity \cite{burek2017fiber}.

\begin{figure*}
\begin{center}
		\includegraphics[width=1.75\columnwidth]{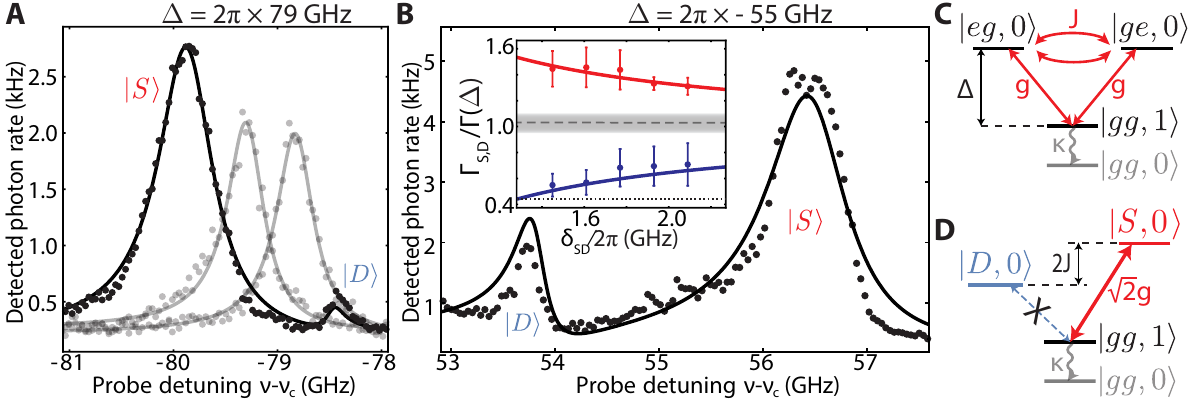}
\end{center}
		\vspace{-0.25cm}
		\caption{
	\textbf{Cavity mediated SiV-SiV interactions.}
	\textbf{(A)} 
	Transmission spectrum of two nearly-resonant SiVs (SiV-SiV detuning $\delta =\,2\pi\times$\,\SI{0.56}{GHz}) at cavity detuning $\Delta =\,2\pi\times$ \SI{79}{GHz}.
	When both SiVs are simultaneously coupled to the cavity, superradiant (bright) $\ket{S}$ and subradiant (dark) $\ket{D}$ collective states are formed (black).
	Individual spectra of non-interacting SiVs are shown in gray.
	\textbf{(B)}
	Transmission spectrum of the same SiVs at an opposite cavity detuning $\Delta =\,2\pi\times$\,\SI{-55}{GHz} and $\delta =\,2\pi\times$\,\SI{2}{GHz}.
	Inset: ratio of $\ket{S}$ (red) and $\ket{D}$ (blue) state linewidths to the single-SiV linewidth at $\Delta =\,2\pi\times$ \SI{79}{GHz} as a function of $\delta_{SD}$.
	The resonance frequencies of these SiVs slowly drift due to spectral diffusion\cite{evans2016narrow}, allowing us to measure the linewidths of the superradiant and subradiant states at different $\delta_{SD}$.
	The gray dashed line and shaded region are the average and standard deviation of the single-SiV linewidths.
	The dotted line corresponds to the SiV linewidth $\gamma$ without Purcell enhancement, demonstrating that the subradiant state at minimum $\delta_{SD}$ is almost completely decoupled from the cavity mode.
	Solid lines in (A) and (B) are predictions based on independently-measured SiV parameters \cite{SOM}.
	\textbf{(C)}
	Energy diagram of two SiV centers coupled to a cavity mode. Two SiVs are detuned from the cavity by $\Delta$, are resonant with each other and are each coupled to the cavity with single-photon Rabi frequency $g$. They interact via exchange of a 
	cavity photon at a rate $J = g^2/\Delta$, forming collective $\ket{S}$
	and $\ket{D}$
	eigenstates \textbf{(D)}.
		}
	\label{fig:2SiV}
\end{figure*}

As is evident from \fignumnp{1C}, SiV centers are subject to inhomogeneous broadening, resulting predominantly from local strain within the nanophotonic device \cite{evans2016narrow}.
This broadening is significantly smaller than that of other solid-state emitters when compared to the lifetime-limited emitter linewidths \cite{lodahl2015interfacing,laucht2010mutual,kim2011strong}. In fact, the resonance frequencies of some SiV centers within the same devices are nearly identical.
To study the cavity-mediated interaction between SiV centers, we focus on a pair of such nearly-resonant SiV centers
(SiV-SiV detuning $\delta=2\pi\times$ \SI{0.6}{GHz})
that are coupled to the cavity in the dispersive regime, that is, with large SiV-cavity detuning ($\Delta = 2\pi\times$ \SI{79}{GHz} $> \kappa$, \fignumnp{2A}).
To identify resonances associated with individual SiV centers, we selectively ionize either SiV into an optically-inactive charge state
by applying a resonant laser field at powers orders-of-magnitude higher than those used to probe the system \cite{SOM}.
This allows us to measure the transmission spectrum for each of the two SiV centers
individually, while holding the other experimental parameters (such as $\Delta$) fixed (\fignumnp{2A}, gray data).

When both SiV centers are in the optically-active charge state, we observe a significantly increased splitting of the transmission resonances.
The new resonances (\fignumnp{2A}, black data) also display different transmission amplitudes compared with the single-SiV resonances, and are labeled as bright ($\ket{S}$) and dark ($\ket{D}$) states, respectively.
The amplitudes and linewidths of $\ket{S}$ ($\ket{D}$) are enhanced (suppressed) compared to those of the individual SiV centers (\fignumnp{2B}, inset).
At a cavity detuning of the opposite sign ($\Delta = 2\pi\times$ \SI{-55}{GHz}), we observe that the sign of the energy splitting $\delta_{SD}$ between $\ket{S}$ and $\ket{D}$ is reversed \fignum{2B}, indicating that the cavity resonance affects $\delta_{SD}$.

\begin{figure}
\begin{center}
		\includegraphics[width=\columnwidth]{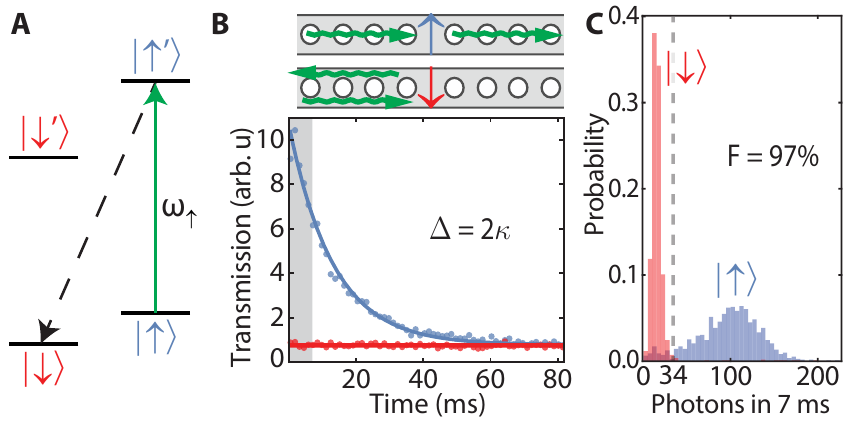}
\end{center}
		\vspace{-0.25cm}
		\caption{
			\textbf{Cavity-assisted spin initialization and readout.}
			\textbf{(A)}
			Simplified level structure of the SiV in a magnetic field. An optical transition at frequency $\omega_\uparrow$ (green arrow) is used to initialize the SiV spin into $\ket{\downarrow}$ by optical pumping via a spin-flipping transition (dashed line). 
			Conversely, pumping at frequency $\omega_\downarrow$ (not shown) initializes the spin into $\ket{\uparrow}$.
			\textbf{(B)}
			Spin-dependent optical switching in the dispersive regime.
			State $\ket{\downarrow}$ is not coupled to the probe field at frequency $\omega_\uparrow$ which is therefore reflected (red).
			Initialization into $\ket{\uparrow}$ results in transmission of the probe field
			(blue).
			\textbf{(C)}
			Photon number distributions for transmission in the dispersive regime for initialization into $\ket{\uparrow}$ (blue) and $\ket{\downarrow}$ (red). The distributions are well-resolved (mean $n_\uparrow=96$, $n_\downarrow=16$) in a \SI{7}{ms} readout window (gray region in \textbf{(C)}), demonstrating single-shot spin-state readout with 97\% fidelity.
		}
	\label{fig:spin}
\end{figure}

To understand these observations, we consider  
the system of two SiV centers coupled to a cavity mode, described by the Hamiltonian \cite{zheng2000efficient,majer2007coupling}:
\begin{equation*}
\label{eqn:HTC}
\begin{split}
\hat{H}/\hbar=&\ \omega_c \hat{a}^\dag \hat{a} + 
					 \omega_1 \hat{\sigma}_1^\dag \hat{\sigma}_1 +
					 \omega_2 \hat{\sigma}_2^\dag \hat{\sigma}_2\ + \\
					 \quad& \hat{a}^\dag\left(g_1\hat{\sigma}_1+g_2\hat{\sigma}_2\right) +
					 \hat{a}(g_1^*\hat{\sigma}_1^\dag +g_2^*\hat{\sigma}_2^\dag)
\end{split}
\end{equation*}
where $\omega_c$ and $\omega_i$ are the frequencies of the cavity and $i$\textsuperscript{th} SiV center and the operators $\hat{a}$ and $\hat{\sigma}_i$ are the photon annihilation and $i$\textsuperscript{th} SiV center's electronic state lowering operators, respectively. 
The coherent evolution of $\hat{H}$ is modified by the decay of the cavity field ($\kappa$) and SiV decay and decoherence ($\gamma$) \cite{SOM}.
In the dispersive regime, $\hat{H}$ yields an effective Hamiltonian for two resonant ($\delta=0$) SiV centers \cite{zheng2000efficient,majer2007coupling}:
\begin{math}
\label{eqn:Heff}
\hat{H}_{\mathrm{eff}}/\hbar=
J\left(\hat{\sigma}_1\hat{\sigma}_2^\dag+\hat{\sigma}_1^\dag\hat{\sigma}_2\right)
\end{math}
where $J=\frac{g^2}{\Delta}$ 
(in our system, $g_1\approx g_2\equiv g$).
Thus, the two SiV centers interact at a rate $J$ via a flip-flop interaction that is mediated by the exchange of cavity photons \fignum{2C}.
This interaction hybridizes the two resonant SiV centers, forming collective eigenstates from the atomic ground ($\ket{g}$) and excited ($\ket{e}$) states which, for $\delta=0$, take the form
$\ket{S} = \frac{1}{\sqrt{2}}(\ket{eg}+\ket{ge})$ and $\ket{D} = \frac{1}{\sqrt{2}}(\ket{eg}-\ket{ge})$ 
and are split in energy by $2J$ \fignum{2D} \cite{majer2007coupling}.
The symmetric superradiant state $\ket{S}$  has an enhanced coupling to the cavity of $\sqrt{2}g$ (making it ``bright'' in transmission) and an energy shift (AC Stark shift) of $2J=2\frac{g^2}{\Delta}$, whereas the antisymmetric combination $\ket{D}$ is completely decoupled from the cavity (``dark'' in transmission) and has a vanishing energy shift \cite{kim2011strong,majer2007coupling}.
The energy shift of state $\ket{S}$ is always away from the cavity mode, explaining the reversed energy difference $\delta_{SD}$ between states $\ket{S}$ and $\ket{D}$ upon changing the sign of $\Delta$ \fignum{2B}.
By comparing the experimental data in \fignumnp{2} to theoretical predictions that account for the finite two-SiV detuning (\fignumnp{2}, solid curves), we extract the SiV-SiV interaction strength $J = 2\pi\times$ \SI{0.6}{GHz}. 
This model uses independently-measured SiV-cavity parameters; the only free parameters correspond to the background field and the amplitude of the signal.
The energy splitting $\delta_{SD}$ (which is at least $2J$) is larger than the measured linewidths (for a single SiV, $\Gamma(\Delta=79\,\mathrm{GHz}) = 2\pi\times$ \SI{0.4}{GHz}), allowing us to spectrally resolve these states.

We next use the SiV center's long-lived electronic spin degree of freedom \cite{hepp2014electronic} to deterministically control both the SiV-cavity transmission and the two-SiV interaction.
We access the spin by
applying a magnetic field
to lift the degeneracy of the spin sublevels in the lower-energy orbital branches of the ground (spin states $\ket{\uparrow}$ and $\ket{\downarrow}$) and optically-excited ($\ket{\uparrow'}$ and $\ket{\downarrow'}$) states.
The Zeeman shifts are different for each orbital state and depend on both the magnitude and orientation of the field with respect to the SiV center's symmetry axis, yielding spectrally-distinguishable spin-selective optical transitions at frequencies $\omega_\uparrow$ and $\omega_\downarrow$ \fignum{3A}.
In general, the splitting between these frequencies is maximized for large off-axis magnetic fields \cite{hepp2014electronic}.
Also, in the presence of any off-axis magnetic field component, the optical transitions are not perfectly cycling, 
allowing us to initialize  
the SiV center
into $\ket{\uparrow}$ by pumping at $\omega_\downarrow$ and vice versa \cite{rogers2014all,pingault2014all}.
The use of spin-selective transitions coupled to the cavity mode 
directly enables high-contrast spin-dependent modulation of the cavity transmission.

\begin{figure*}
\begin{center}
		\includegraphics[width=1.75\columnwidth]{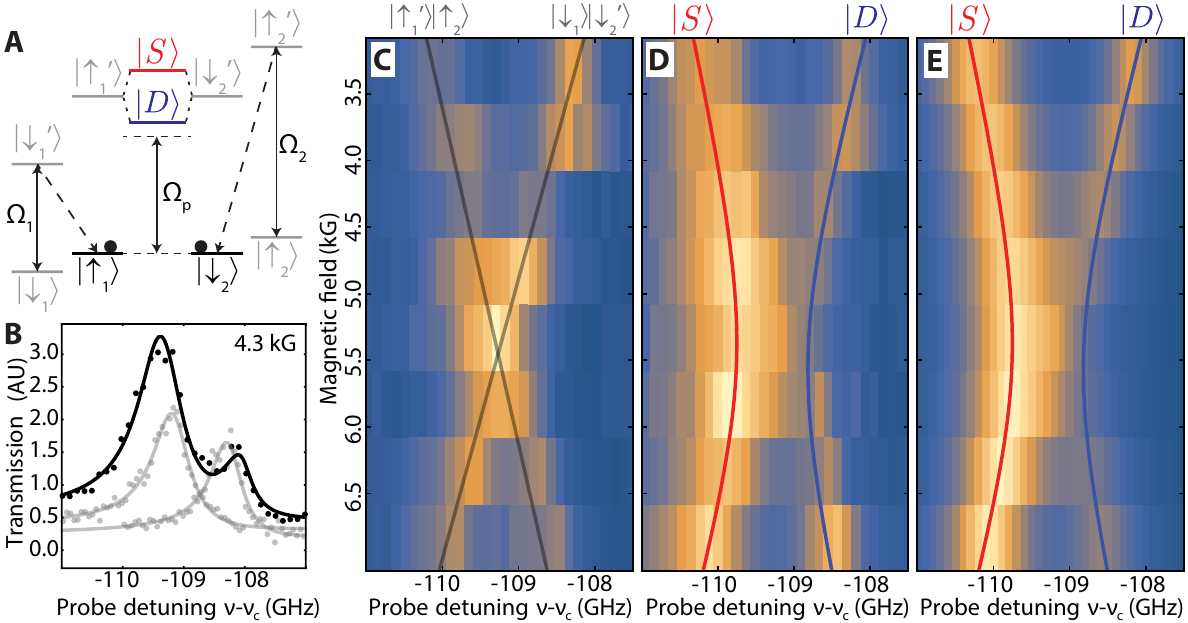}
\end{center}
		\vspace{-0.25cm}
		\caption{
	\textbf{Deterministic control of photon-mediated interactions via SiV spin states}
	\textbf{(A)} 
	Energy diagram of two SiV centers in a magnetic field. The $\ket{\uparrow_1}\rightarrow\ket{\uparrow_1'}$ and $\ket{\downarrow_2}\rightarrow\ket{\downarrow_2'}$ transitions are tuned in and out of resonance by sweeping a magnetic field.
	The spins are pumped with control fields $\Omega_1$ and $\Omega_2$, and the cavity transmission spectrum is measured by varying the frequency of a weak probe field $\Omega_p$.
	When the spins are initialized in $\ket{\uparrow_1}\ket{\downarrow_2}$, collective super- ($\ket{S}$) and subradiant ($\ket{D}$) states are formed.
	\textbf{(B)}
	Spin-dependent transmission spectra at a magnetic field of \SI{4.3}{kG}.
	Spectra of single SiVs in the noninteracting combinations of spin states are shown in gray.
	The spectrum of two interacting SiVs (black) demonstrates formation of $\ket{S}$ and $\ket{D}$.
	\textbf{(C)}
	Composite spectrum of the non-interacting system at different two-SiV detunings.
	The solid lines are the fitted single-SiV energies of $\ket{\uparrow_1'}\ket{\uparrow_2}$ and $\ket{\downarrow_1}\ket{\downarrow_2'}$ as a function of magnetic field.
	\textbf{(D)}
	An avoided crossing is visible in cavity transmission when the spins are prepared in the interacting state $\ket{\uparrow_1}\ket{\downarrow_2}$.
	\textbf{(E)} Predicted transmission spectrum for independently measured SiV-cavity parameters. The red and blue solid lines in (D) and (E) are predicted energies of $\ket{S}$ and $\ket{D}$ based on these parameters \cite{SOM}.
		}
	\label{fig:anticrossing}
\end{figure*}

We demonstrate this effect by focusing on a single SiV center in the dispersive regime ($\Delta\sim2\kappa$).
Here, the optical transition linewidth is narrow, allowing us to resolve these transitions in a $\SI{9}{kG}$ magnetic field almost perfectly aligned with the SiV center's symmetry axis
where the transitions are
highly cycling (branching fraction $\sim1-10^{-4}$) \cite{sukachev2017silicon}. We initialize the spin in either $\ket{\downarrow}$ or $\ket{\uparrow}$ via optical pumping and probe the system in transmission.
When the spin is in $\ket{\downarrow}$, the interaction with the probe field at $\omega_\uparrow$ is negligible and the probe is reflected by the detuned cavity (\fignumnp{3B}, red curve). When the spin is in $\ket{\uparrow}$, single photons at frequency $\omega_\uparrow$ are transmitted via the SiV-like dressed state
(blue curve) for a time (\SI{12}{ms}) determined by the cyclicity of the optical transition \cite{sukachev2017silicon}.
We construct a histogram of photons detected in a \SI{7}{ms} window when the spin is initialized in $\ket{\uparrow}$ (red) and $\ket{\downarrow}$ (blue) \fignum{3D}. The photon count distributions for the two spin states are well-resolved, allowing us to determine the spin state in a single shot with $97 \%$ fidelity \cite{SOM}.
We also perform this experiment in the resonant-cavity regime and observe spin-dependent switching of the transmission with a maximum contrast of 80\% \cite{SOM}.

The combination of spin control, high-cooperativity coupling and a relatively small inhomogeneous distribution of SiVs enables controllable optically-mediated interactions between multiple SiV centers.
We focus on two SiV centers
(SiV 1 and 2)
in the dispersive regime ($\Delta = 2\pi\times$ \SI{109}{GHz}) with cavity QED parameters 
$\{g_1\approx g_2,\kappa,\gamma_1\approx\gamma_2\}=2\pi\times\{7.3,39,0.5\}$~\si{GHz} ($C\approx11$)
that are initially detuned from one another by $\delta =2\pi\times$ \SI{5}{GHz} \cite{SOM}. 
We sweep the magnitude of a magnetic field oriented 
almost orthogonal to the SiV symmetry axis and tune transitions $\ket{\uparrow_1} \rightarrow \ket{\uparrow_1'}$ and $\ket{\downarrow_2} \rightarrow \ket{\downarrow_2'}$ (which have opposite Zeeman shifts) in and out of resonance \fignum{4A}.
At each magnetic field, we use a continuous field $\Omega_1$ or $\Omega_2$ to optically pump either SiV 1 or SiV 2 into the spin state resonant with a weak probe field $\Omega_p$
and measure the transmission spectrum of the system.
This allows us to perform control measurements where only one spin is addressed by $\Omega_p$ at a time (gray data in \fignumnp{4B}). 
The single-spin transmission spectra at each field are summed to form a composite spectrum of the non-interacting two-SiV system \fignum{4C}, which displays an energy level crossing of the two SiV transitions, characteristic of non-interacting systems. 

We next perform the measurements in the interacting regime by preparing the spins in the state $\ket{\uparrow_1}\ket{\downarrow_2}$
by applying both fields $\Omega_1$ and $\Omega_2$.
The two-SiV transmission spectrum demonstrates the formation of superradiant and subradiant states (\fignumnp{4B}, black) that exist only for this combination of spin states.
Transmission spectra as a function of the applied magnetic field are shown in Figure {\bf 4D}, demonstrating an avoided crossing arising from spin-dependent interactions between the two SiV centers 
\cite{majer2007coupling}.
These experimental observations are in excellent agreement with theoretical predictions based on 
the independently-measured SiV-cavity parameters 
\fignum{4E}.
Similar observations were reproduced in a separate device on the same chip \cite{SOM}.

The optically-mediated interaction between quantum emitters observed here could be used 
to realize key quantum information protocols \cite{imamoglu1999quantum}, including
cavity-assisted entanglement generation \cite{zheng2000efficient,kastoryano2011dissipative}, efficient Bell-state measurements \cite{waks2006dipole,borregaard2015long} and 
robust photon-mediated gates between emitters in distant cavities \cite{cirac1997quantum,monroe2014large}.
To implement these schemes with high fidelity, qubits should be encoded in long-lived SiV electronic spin states.
Recent work has already demonstrated that the SiV spin can be used as a long-lived quantum memory \cite{sukachev2017silicon} that can be coherently manipulated with both microwave \cite{sukachev2017silicon,pingault2017coherent} and optical fields \cite{becker2018all}.
The fidelity associated with cavity-mediated quantum operations between such qubit states is limited by the cooperativity \cite{borregaard2015long,kastoryano2011dissipative}.
While the cooperativity $C \sim 20$ achieved in this work is among the largest demonstrated in the optical domain, it can be further improved by at least two orders of magnitude by increasing the cavity $Q/V$ and by reducing sources of spectral diffusion which limit $\gamma$.
Alternatively, the cooperativity could be enhanced by using different quantum emitters, such as the GeV \cite{iwasaki2015germanium} or SnV \cite{iwasaki2017tin}
centers in diamond, which feature higher quantum efficiencies \cite{bhaskar2017quantum}. 
Near-unity fidelities can also be achieved with existing cooperativities using recently-proposed heralded schemes where gate errors can be suppressed via error detection with an auxilary qubit \cite{borregaard2015long}. Furthermore, our system could be used to efficiently generate
non-classical states of light \cite{economou2010optically}, which are useful in, for example, measurement-based quantum computing.
On-chip scalability and GHz-level photon bandwidths
make our system particularly well-suited for applications in quantum networking, 
paving the way for implementation of efficient quantum repeaters \cite{muralidharan2016optimal,borregaard2015long}
and distributed quantum computing \cite{monroe2014large,imamoglu1999quantum}.

\section*{Acknowledgments}
\begin{acknowledgments}
We thank D.\,Twitchen and M.\,Markham from Element Six Inc.\ for substrates and J.\,Borregaard and K.\,De Greve for useful discussions. 
D.\,Perry assisted with ion implantation.
Financial support was provided by the NSF, the CUA,
the DoD/ARO DURIP program,
the AFOSR MURI, 
the ONR MURI,
the ARL, the Vannevar Bush Faculty Fellowship program, the DoD NDSEG (M.\,K.\,B.), and the NSF GRFP (B.\,M. and G.\,Z.).
Devices were fabricated at the Harvard CNS supported under NSF award ECCS-1541959.
Ion implantation was performed at Sandia National Laboratories through the Center for Integrated Nanotechnologies, an Office of Science facility operated for the DOE (contract DE-NA-0003525) by Sandia Corporation, a Honeywell subsidiary.
The views expressed here do not necessarily represent the views of the U.\,S.\ Department of Energy (DOE) or the U.\,S.\ Government.
\end{acknowledgments}

\bibliographystyle{PRAppl_Denis}
\bibliography{SiVbib_arxiv}

\end{document}